\documentclass[12pt]{article}
\usepackage{graphics}
\begin{document}

\begin{center}
{\Large HD 112914 : A nearby one solar mass binary system}

\vspace{5mm}

{\it Henri M.J. Boffin$^1$} and {\it Dimitri Pourbaix$^2$}
\end{center}

\vspace{5mm}
{\it 
$^1$ Royal Observatory of Belgium, Avenue circulaire 3, 1180 Brussels

$^2$ Research Associate FNRS, Institut d'Astronomie et d'Astrophysique, Universit\'e Libre de Bruxelles, 
C.P. 226, Bd de la Plaine, 1050 Brussels
}

\vspace{1cm}

{\large In paper 167 of his serie published in The Observatory, Griffin$^1$ presented
the spectroscopic orbit of HD 112914, a late main sequence star. He also noticed
that this star, also known as HIP 63406, was one of the few for which the DMSA/O Annex of the Hipparcos 
Catalogue$^2$ derived an orbit prior to any spectroscopic one. Albeit in agreement
with each others, the astrometric orbit was however determined with rather large
uncertainties. Here, we have reanalysed the Hipparcos Intermediate Astrometric Data
(IAD) using Griffin'spectroscopic orbit to obtain a much more precise astrometric orbit. 
Several parameters of the HD 112914 system are now well constrained.
}

\vspace{1cm}

We have fitted the Hipparcos IAD with two 
distinct sets of orbital parameters and assessed the reliability of the
orbit from the agreement between the two solutions$^3$.
Let us briefly summarize the two approaches, based respectively on the Campbell method and on the Thiele-Innes constants, keeping in mind that in both cases,
the eccentricity, orbital period and periastron time are assumed from the spectroscopic orbit.  On the one hand, the four remaining parameters, i.e. the semi-major axis of the photocentric orbit ($a_0$), the inclination ($i$), the latitude
of the ascending node ($\Omega$) and the argument of the periastron ($\omega$) are fitted through their Thiele-Innes constant combination.  On the other hand, two more parameters are assumed from the spectroscopic orbit, namely the amplitude of the radial velocity curve and $\omega$.  Here, only two parameters of the photocentric orbit ($i$ and $\Omega$) are thus derived.  Although that double-fit approach was designed in the context of the astrometric orbits of the extrasolar planets, it can be applied to spectroscopic binaries, especially single-lined$^4$.
In the case of HD 112914, 
the fit of the IAD is significantly improved with an orbital model 
and the fitted orbital parameters (Thiele-Innes' constants) are significantly non zero.
The consistency of the Thiele-Innes
solution and the Campbell/spectroscopic one is verified at the 99.85 \% level, 
indicating an extraordinarily good agreement between the spectroscopic and
the astrometric solution.

\begin{table}
\begin{center}
{\bf Table 1 \\ Astrometric orbit of HD 112914}\\
\vspace{1mm}
\begin{tabular}{ll}
parallax (mas)   & 40.02 $\pm$ 1.13\\
inclination (deg )& 80.9 $\pm$ 5.2 \\
$a_0$ (mas) & 14.1 $\pm$ 1.35\\
$\Omega$ (deg) & 102      \\
\end{tabular}
\end{center}
\end{table}

The astrometric parameters we derive for HD 112914 taking into account the
spectroscopic orbit are given in Table 1. Our parallax is slightly smaller than 
the original Hipparcos one but both agree within the error bars. The inclination 
we obtain is also very similar although with a slightly improved accuracy. The
motion of the photocentre - $a_o$ - however has been reduced and is now twice
as precisely determined as before. 

\begin{figure}[t]
\begin{center}
\resizebox{0.75\hsize}{!}{\includegraphics{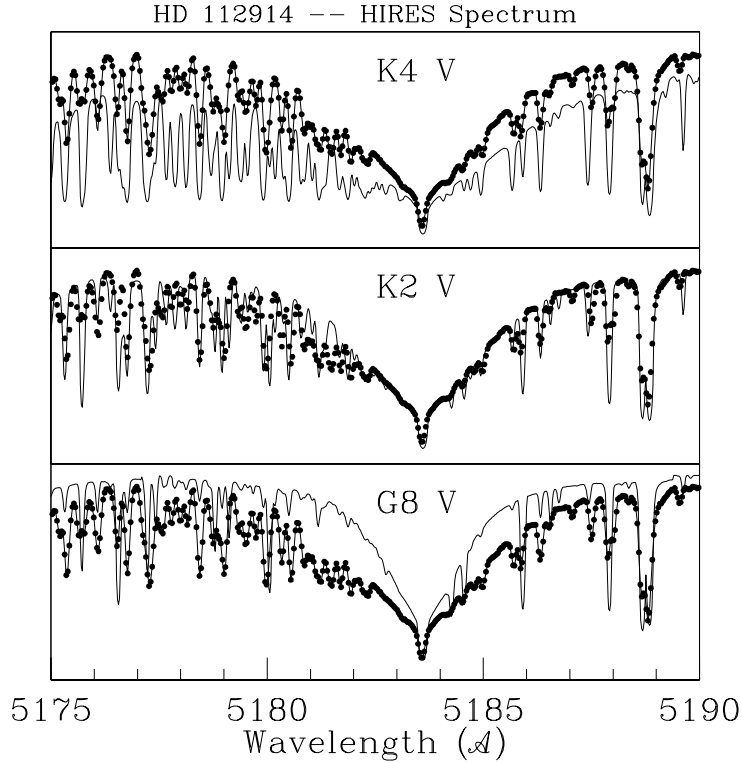}}
\end{center}
\caption{Part of the Keck HIRES spectrum of HD 112914 (heavy dots) is compared with synthetic
spectra computed using the Spectrum package and three different Kurucz models for main sequence
stars of spectral type G8, K2 and K4 V.}
\end{figure}

We can now make use of all our informations on HD 112914 to try to constrain the
components of the systems. 
It is of interest to note that from the spectroscopic orbit, the mass function is
known with an accuracy of 4 \%. This combined with the error on the inclination coming from the astrometric orbit, leads to a formal error for the ratio
$m_2^3/(m_1+m_2)^2$ of 6 \%. For a given primary mass, $m_1$, the companion mass, $m_2$, is thus
known with very great accuracy.

Presented in the Simbad database as a G9V high proper-motion star, the colour
of HD 112914 seems to contradict the first assertion. Its B-V index of 0.94 
is indeed closer to a K2-K3 V than a G9 V. Moreover, using the Hipparcos
parallax and noting that at the position of HD 112914, there is little 
interstellar extinction, the +6$^{\rm m}$.62  of its absolute visual magnitude
is also more adequate for a K3 V than for a G9 V star. 
The Yoss-Griffin$^5$ survey gave the DDO-based type of K3 V and the absolute magnitude of +6$^{\rm m}$.5.

We were able to secure a Keck HIRES echelle spectrum of HD 112914 which covers the wavelength range from 
3890 to 6190 \AA. We have used the Spectrum package$^6$ to compare 
the reduced spectrum with synthetic spectra using Kurucz models of main sequence stars with effective
temperature in the range 4500-5500 K. Our analysis clearly shows that HD 112914 is a K2 or K3 V star with 
an effective temperature of 5000$\pm 100$ K. A small part of the spectrum is compared in figure 1 
with the synthetic spectrum of main sequence stars, with spectral types G8, K2 and K4 V.

As the  DDO photometry and our high-resolution spectrum reveal an apparently solar metallicity star, we
can deduce the mass of the primary using the analytical evolutionary 
tracks of Hurley et al.$^7$. We find a value of 0.75 M$_\odot$ if the star has an age 
of 15 Gyr to a limiting 0.81 M$_\odot$ if the star just formed.
With the above quoted value for $m_1$, we derive $m_2=0.23 \pm 0.01$  M$_\odot$,
i.e. a mass ratio of $0.29 \pm 0.01$.
The total mass of the system is thus, rather amazingly, $1.00 \pm 0.03$ M$_\odot$. 
With such a low mass, the secondary does not contribute any light to the 
system and the motion of the photocentre represents the orbital motion of the
primary. This is indeed verified as the $a_1$ value obtained from the 
spectroscopic orbit (52.5 Gm) is totally compatible with the $a_o$ value
we obtain (0.35 A.U.). 

Using the Delfosse et al.$^8$ mass-luminosity relation for low-mass stars, we
can compute that the secondary has an absolute visual magnitude of 12.68 and hence the magnitude difference between the two objects is almost exactly 6 magnitudes. 

HD 112914 is thus a K2-3 V star, 25 pc away from the Sun, of mass close to 0.78  M$_\odot$. Its 0.23 M$_\odot$ companion orbits
with a semi-major axis of 1.54 A.U.
This results is a nice example of the wealth of information still available in the Hipparcos
Catalogue when its data will be reanalysed more thoroughly along new 
spectroscopic orbits. It is also a nice premice of the way 
other astrometric missions, and especially GAIA, will revolutionise our 
knowledge of binary stars.

\vspace{5mm}
\noindent {\it \small This research was supported in part by ESA/PRODEX 14847/00/NL/SFe(IC) and C15152/01/NL/SFe(IC). It is a pleasure to 
thank Paul Butler, Geoff Marcy and Jason Wright for kindly taking a spectrum of HD 112914 with the Keck. We also thank Richard O. Gray 
for making his Spectrum package available.}

\vspace{5mm}

\noindent {\it \large References}

\noindent (1) R.F. Griffin, {\it The Observatory}, {\bf 122}, 329, 2002 (Paper 167).

\noindent (2) ESA, 1997, The Hipparcos and Tycho Catalogues, ESA SP-120

\noindent (3) D. Pourbaix \& F. Arenou, {\it A\&A}, {\bf 372}, 935, 2001.

\noindent (4) D. Pourbaix \& H.M.J. Boffin, {\it A\&A}, 398, 1163, 2003.

\noindent (5) K.M. Yoss \& R.F. Griffin, {\it JA\&A}, {\bf 18}, 161, 1997.

\noindent (6) R.O. Gray, 2001, http://www.phys.appstate.edu/spectrum/spectrum.html

\noindent (7) J.R. Hurley, O.R. Pols, C.A. Tout, {\it MNRAS}, {\bf 315}, 543, 2000.

\noindent (8) X. Delfosse, T. Forveille, D. Ségransan, J.-L. Beuzit, S. Udry, C. Perrier \& M. Mayor, {\it A\&A}, {\bf 364}, 217, 2000.
\end{document}